\documentclass[final,3p,times,english]{elsarticle}

\usepackage{graphicx}
\usepackage{subfigure}
\usepackage{tikz}
\usepackage{pgfplots}
\usepackage{hyperref}
\hypersetup{
  colorlinks,
  citecolor=Violet,
  linkcolor=Red,
  urlcolor=Blue}
\usepackage{amssymb}
\usepackage{amsmath}
\usepackage{multirow}
\usepackage{makecell}
 \usepackage{kpfonts}
 \usepackage{cleveref}
\usepackage{lineno}

\usepackage{multirow}
\usepackage{threeparttable}
\usepackage{tabularx}
\usepackage{color}
\usepackage{soul}
\usepackage{multirow}
\usepackage{booktabs}
\usepackage{array}
\usepackage{csquotes}

\usepackage[bold]{hhtensor}

\biboptions{numbers,comma,square,sort&compress}

\usepackage[T1]{fontenc}
\usepackage{charter}
\usepackage[expert]{mathdesign}

\usepackage[utf8]{inputenc}

\usepackage{units}
\usepackage{babel}
\usepackage{sublabel}
\usepackage{tikz}
\usetikzlibrary{arrows,decorations.markings,backgrounds,positioning,fit,shapes,patterns,fadings}
\usepackage{tikz-3dplot}
\usepackage{mathabx}

\usepackage{miller}
\usepackage{csquotes}
\usepackage{adjustbox}

\usepackage{lscape}

\usepackage{algorithm}
\usepackage[noend]{algpseudocode}
\newcommand{\pluseq}{\mathrel{+}=}

\begin{document}
\newcommand{\etal}{{\it et al.}}

\begin{frontmatter}



\title{First-principles study of the energetics and the local chemical ordering of tungsten-based alloys}

\author[add1]{Yichen Qian}
\author[add2]{Mark R. Gilbert}
\author[add3]{Lucile Dezerald}
\author[add2]{Duc Nguyen-Manh}
\author[add1]{David Cereceda\corref{cor1}}
\ead{david.cereceda@villanova.edu}
\cortext[cor1]{Corresponding author}
\address[add1]{Department of Mechanical Engineering, Villanova University, Villanova, PA 19085, USA}
\address[add2]{United Kingdom Atomic Energy Authority, Culham Centre For fusion Energy, Culham Science Centre, Abingdon, Oxon, OX14 3DB, UK}
\address[add3]{Department of Materials Science and Engineering, Institut Jean Lamour, Universite de Lorraine, F-54011 Nancy, France}

\begin{abstract}
Tungsten (W) is considered a leading candidate for structural and functional materials in future fusion energy devices. The most attractive properties of tungsten for magnetic and inertial fusion energy reactors are its high melting point, high thermal conductivity, low sputtering yield, and low long-term disposal radioactive footprint. However, tungsten also presents a very low fracture toughness, primarily associated with inter-granular failure and bulk plasticity, limiting its applications. 
In recent years, several families of tungsten-based alloys have been explored to overcome the aforementioned limitations of pure tungsten. These might include tungsten-based high-entropy alloys (W-HEAs) and the so-called tungsten-based \textquote{smart alloys}. 
In this work, we present a computational approach that uses first-principles DFT electronic structure calculations to understand the effect of the chemical environment on tungsten-based alloys. In particular, we compared the Special Quasi-random Structure (SQS) and the DFT-coupled Monte Carlo (MC-DFT) methods when investigating the short-range order and elastic properties of two equimolar WCrTaTi and WVTaTi W-HEAs, and two WCrTi and WCrY W-SAs.
We found that structures after MC-DFT calculation have lower cohesive energies than SQS structures, with shorter lattice constants, and large elastic properties values. Furthermore, distinct element segregation was studied by calculating the SRO parameter and radius distribution function. The total density of states suggested that the existence of SRO could improve the stability of the structure.

\end{abstract}


\end{frontmatter}
 


\section{Introduction} \label{sec_Intro}


Tungsten (W) stands out as a leading candidate for plasma-facing applications in magnetic fusion energy (MFE) devices due to its high melting point, excellent thermal conductivity, low sputtering yield, and low long-term radioactive waste impact. However, despite these advantages, tungsten's very low fracture toughness, leading to brittle trans- and inter-granular failure, significantly limits its operating temperature range. \cite{Zinkle_GhoniemJNM11}.

In recent years, several families of tungsten-based alloys have been explored to overcome the aforementioned limitations of pure tungsten. 
High-entropy alloys (HEAs) are a promising class of materials with remarkable properties \cite{tsai2014high,li2016metastable,george2019high,murty2019high}. They were originally conceived in the early 2000s as a blend of five or more elements with individual concentrations between 5 and 35 atom percent \cite{yeh2004nanostructured}. Interestingly, the composition stability of different HEAs phases is strongly correlated with the valence electron concentration from electronic structure analysis \cite{leong2017}.  
In particular, tungsten-based HEAs (W-HEAs) have shown superior mechanical properties at high temperatures, a superior melting point (above 2873 K), enhanced radiation resistance to heavy ion irradiation, and negligible radiation hardening when compared to pure tungsten \cite{waseem2017powder,zou2015ultrastrong,senkov2018development,el2021helium,el2019outstanding,senkov2011mechanical}. 
%
%
Another attractive option are the so-called tungsten-based \textquote{SMART alloys} (W-SAs), where SMART stands for {\bf S}elf-passivating {\bf M}etal {\bf A}lloys with {\bf R}educed {\bf T}hermo-oxidation, that can adapt their behavior to the environment \cite{litnovsky2021,sobieraj2021,litnovsky2017smart,klein2020tungsten,koch2007self}.
For example, in the event of a loss-of-coolant accident (LOCA) combined with an air ingress, W-SAs containing small amounts of Ti or Y have demonstrated the capability to create stable oxides that prevent their mobilization into the atmosphere. 


Despite the numerous efforts in the literature to investigate the effects of alloying elements on various properties of W-based materials such as phase phase stability \cite{Muzyk_2011,muzyk2013first,wei2014first,2018_chaoming,huang2017mechanism}, elastic properties \cite{Muzyk_2011,muzyk2013first,wei2014first,jiang2016mechanical,li2016ab,hu2016effects,2018_chaoming}, ideal tensile strength \cite{2013_GIUSEPPONI,2018_chaoming}, ductility \cite{QIAN_2018}, radiation defects \cite{muzyk2013first,el2019outstanding,Duc2021PRM}, point defects \cite{Muzyk_2011,suzudo2014stability,hossain2014stress,gharaee2016role,giusepponi2015effects,setyawan2017ab,Duc2021PRM}, screw dislocation structures \cite{2010_Romaner,li2012dislocation}, grain boundaries \cite{wu2016first}, transmutation effects \cite{qian2021using,qian2023ab},  etc.,
to the best of our knowledge, there is still a lack of understanding of the local chemical ordering of tungsten-based PFMs.

In this paper, we present a comparison of three different ab initio approaches to describe the disordered alloys systems, special focus on High entropy alloys and smart alloys:(1) Virtual crystal approximation (VCA) ; (2) the special quasi-random structure (SQS); (3) and Monte Carlo couple Density functional theory (MC-DFT). The virtual crystal approximation is computationally cheap among the three approaches because only a conventional bcc structure was used to represent the arbitrary composition of alloys. The SQS method captures the local lattice distortion of alloying elements compared with VCA method, however, the chemical short-range order (CSRO) was missed since it attempts to create a perfect random structure. The CSRO and local lattice distortion could be retained by MC-DFT method, but it requires extreme expensive computational resources.  
%


Our paper is organized as follows.  
After this introduction, we provide in Section~\ref{sec_Methods} an overview of the computational methods employed, including the basic idea of Monte Carlo coupled DFT, short-range order parameter,  and calculation methods of elastic properties. The results are given in Section~\ref{sec_Results}, which includes the energy evaluations during the MC-DFT calculation. 
We finalize in Section~\ref{sec_Discussion} with a brief discussion and the conclusions in Section~\ref{sec_Conclusions}.



\section{Computational methods} \label{sec_Methods}
The local changes in the chemical composition of W-based PFMs are investigated by identifying their energetically stable arrangement of chemical elements. 
For this purpose, we computed their total energies using first-principles DFT electronic structure methods with the Vienna Ab initio Simulation Package (VASP) \cite{}. The distribution of metal elements was examined by comparing the random distribution of the Special Quasi-random Structures (SQS) \cite{zunger1990special} and the Minimum Energy Configurations (MEC) obtained from a DFT-coupled Monte Carlo simulation (denoted as MC-DFT hereinafter).

First, we constructed 128-atom bcc supercells of the four W-based alloys in Table \ref{table_w_materials}. This includes two W-based high-entropy alloys (W-HEAs) and two W-based smart alloys (W-SAs). The table shows that the chemical composition of the two W-SAs was adjusted with respect to their reference value, given the number of atoms present in our bulk structures. 

\begin{table}[!hbtp]
\caption{Chemical composition of the four W-based alloys investigated in this work.}\label{table_w_materials}
\vspace{2mm}
\centering
\footnotesize
\begin{tabular}{l ccccc ccc}
& & \multicolumn{6}{c}{Chemical composition (at\%)[No. atoms in the supercell]} \\
\cmidrule{3-8}
& &  W &  Cr  &  Ti &  Y &  Ta &  V   \\
\midrule
W high-entropy alloy (HEA1) & \cite{sobieraj2020chemical} and this work & 25 [32] &  25 [32] & 25 [32] &  - & 25 [32] & -   \\
\midrule
W high-entropy alloy (HEA2) & \cite{sobieraj2020chemical} and this work & 25 [32] &  -  & 25 [32] &  - & 25 [32] & 25 [32] \\
\midrule
\multirow{2}{*}{W smart alloy (SA1)} & \cite{litnovsky2017smart} & 67.16 &  26.98 &  5.86 & -  & -  &  -    \\
& This work & 67.19 [86] &  26.56 [34] &  6.26 [8] & -  & -  &  - \\
\midrule
\multirow{2}{*}{W smart alloy (SA2)} & \cite{klein2020tungsten}  & 67.93 &  31.11  & -  &  0.958 & -  & -   \\
& This work & 67.19 [86]&  31.25 [40] & -  &  1.56 [2] & -  & -   \\
%
\bottomrule
\end{tabular} \\
\end{table}

SQS structures were generated for each material system using ATAT \cite{van_de_Walle_2002,van2013efficient}. 
The convergence criteria to select the SQS structure serving as the starting point of the MC-DFT simulations was based on the Warren-Cowley short-range order (SRO) $\alpha^{ij}_{k}$ \cite{methfessel1989high}, defined as:
\begin{equation}\label{equation_SRO}
    \alpha^{ij}_{k} = 1- \frac{\rho ^{ij}_{k}}{c_{i}}
\end{equation}
where $\rho ^{ij}_{k}$ is the possibility to find an atom $i$ in the \textit{kth} nearest-neighbor shell of atom $j$, and $c_{i}$ is the composition of atom $i$ in the structure. 
In a perfect random alloy, $\alpha^{ij}_{k} = 0$ , as $\rho ^{ij}_{k} = c_{i}$. If $\rho ^{ij}_{k} > c_{i}$, it suggests that $i$ and $j$ atom prefer to form a pair at $k$th nearest-neighbor. And $\rho ^{ij}_{k} < c_{i}$ indicates that repulsed interaction between $i$ and $j$ atom at $k$th nearest-neighbor.
Given the proximity of the first and second nearest neighbors in bcc structures, we calculated the SRO for the first and second nearest-neighbor shells, as well as the  average SRO for these two shells, defined as \cite{sobieraj2020chemical,fernandez2017short,mirebeau2010neutron}:

\begin{equation}\label{equation_SRO_ave}
    \alpha^{ij}_{ave} = \frac{8 \times \alpha^{ij}_{1} + 6 \times \alpha^{ij}_{2}}{14}
\end{equation}

\noindent where $\alpha^{ij}_{1}$ and $\alpha^{ij}_{2}$ are the first- and second-nearest neighbor SRO parameters, respectively.
As SQS are generated, their SRO parameters are calculated. Once a specific SQS structure's SRO = 0, no further SQS structures of that material system are created, and that particular structure is taken as the starting point of the MC-DFT simulations.

%


\begin{center}

\begin{minipage}{10cm}

\begin{algorithm}[H]
\caption{Monte Carlo DFT-based approach to find minimum-energy configurations (MECs)}\label{alg_MCDFT}
\begin{algorithmic}[1]
\State Input: SQS structure with energy $E_0$
\State Let $E_i$ = $E_0$ 
\While{(step $<$ limit)} $\triangleright$ loop over all MC steps
\State Randomly choose two atoms and swap their positions
\State Compute energy of the swapped configuration, \textit{i.e.}, $E_s$
\If{($E_s < E_i$)} 
    \State Accept swapped configuration, \textit{i.e.}, $E_i = E_s$ 
\Else
    \State Compute probability $ \rho = {\rm e}^{-\frac{(E_s - E_i)}{k_B T}}$
    \State Generate random number $\delta \in (0,1)$
    \If{($\rho < \delta$)}
        \State Accept swapped configuration, \textit{i.e.}, $E_i = E_f$ 
    \Else
        \State Reject swapped configuration
    \EndIf
    \State \textbf{end if}
\EndIf 
\State \textbf{end if}
\State step $\pluseq$ 1
\EndWhile\label{euclidendwhile}
\State \textbf{end while}
\State Output: set of MECs with their corresponding energy and state (accepted/rejected). %
\end{algorithmic}
\end{algorithm}

\end{minipage}

\end{center}

The proposed MC-DFT method consists of sampling the phase space to find MECs by randomly swapping the chemical elements between atom locations. During each MC swap, both the bcc structure and the lattice are fixed while the atomic positions are allowed to relax. After every swap, the change in the Hamiltonian $\Delta H_{i\rightarrow s}$ between the initial $i$ and swapped $s$ configurations is calculated from DFT. The following criterion is applied to accept or reject each event: a new swapped state $S$ is always accepted if the total energy decreases during the swap. If the energy increases, the new state is accepted with the probability $\rho = exp (-\frac{\Delta H_{i\rightarrow s}}{k_{B}T})$, where $k_{B}$ is the Boltzmann constant and $T$ is the Monte Carlo temperature.
The process described above is repeated for a given number of MC steps or until a MEC with a specific energy is found. 
For completeness, the MC-DFT method is detailed in Algorithm \ref{alg_MCDFT}.


All DFT calculations were performed on defect-free $4 \times 4 \times 4$ bcc supercells containing 128 atoms, using the Vienna Ab initio Simulation Package (VASP) \cite{kresse1993ab} with projector augmented wave (PAW) pseudo-potentials \cite{blochl1994projector} and the Perdew-Burke-Ernzerhof (PBE) exchange-correlation functional \cite{PBE}. 
Energy calculations within the MC-DFT method employed a plane wave cutoff energy of 300 eV and a $3 \times 3 \times 3$ \textit{k}-point mesh.
For their part, structural relaxations of the MECs found by the end of the MC-DFT runs were performed using an energy cutoff of 350 eV and a denser \textit{k}-point mesh $6 \times 6 \times 6$.
All calculations were performed without spin polarization included as there is no experimental evidence about magnetic properties in the considered W-SAs and W-HEAs containing Cr. Indeed, previous works in the literature have shown that, for the binary W-Cr system, which is important in our study, no magnetic phases are observed from its phase diagram \cite{handbook1992alloy}.

\section{Results}\label{sec_Results}


\subsection{Energetics of W-based alloys}\label{MEC_search}

\begin{figure}[hbtp!]
\centering
\includegraphics[width=1.0\textwidth]{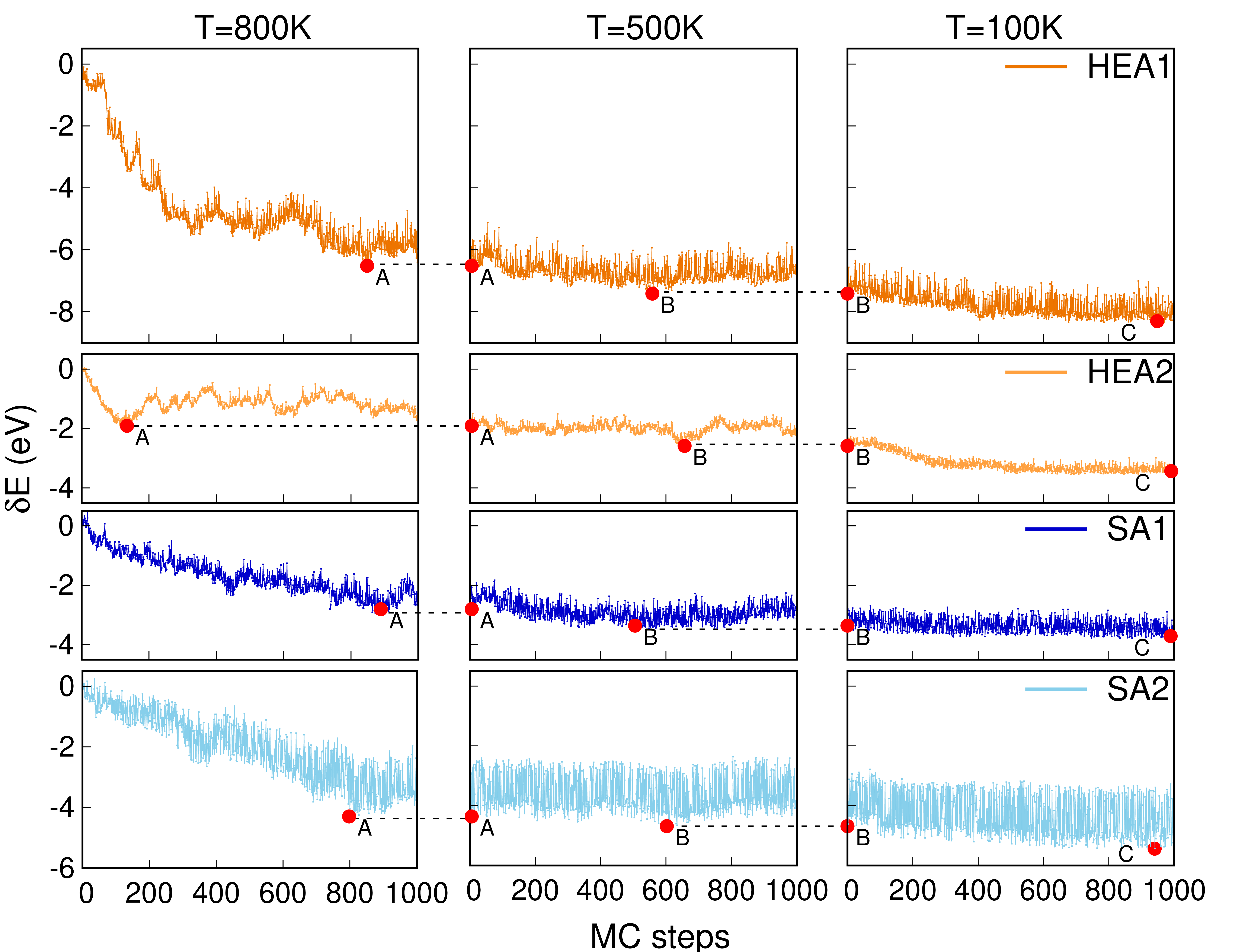}
\caption{Evolution of total energies, with respect to the total energy of the initial SQS configuration, when applying the MC-DFT method sequentially at three different temperatures: 800K, 500K, and 100K.}\label{fig_dE_vs_MCsteps_sequential_All}
\end{figure}

Figure \ref{fig_dE_vs_MCsteps_sequential_All} shows the energy evolution of the four W-based alloys when applying the MC-DFT method described above sequentially at three MC temperatures: 800K, 500K, and 100K. For each alloy, the starting point of this set of simulations is an SQS with SRO = 0 for its first nearest-neighbor (NN) shell.  
Then, the MC-DFT method is executed for 1000 MC steps at each MC temperature, beginning by the highest one. After 1000 MC steps, the MEC that is found at each MC temperature is taken as the starting point of the MC-DFT simulations at the next lower MC temperature. This process is repeated for all MC temperatures. The MEC found at the lowest MC temperature is denoted as the overall MEC configuration of the specific W-based alloy. 
%
The number of swap trials per atom in the MC-DFT runs shown in Fig.~\ref{fig_dE_vs_MCsteps_sequential_All} is limited compared to classical MC simulations. Still, the final MEC obtained for each W-based alloy at the end of the 100K runs has lowered its energy with respect to the initial SQS within a range from -9 eV (HEA1) to -3.5 eV (HEA2). 
In addition to the total change in energy that exists between the initial SQS configuration and the MEC, it is also noticeable the differences that appear when analyzing the energy difference between consecutive MC steps for the four W-based alloys investigated here:

The MC-DFT method was applied sequentially in the results shown in Fig.~\ref{fig_dE_vs_MCsteps_sequential_All} to minimize the changes of the system of getting trapped in local minima during the first MC steps, as higher MC temperatures allow for the acceptance of MC swaps at higher rates, allowing the system to overcome a higher jump in total energy between two consecutive MC steps. 
However, applying the MC-DFT that way increases significantly the computational cost of the simulations. And it also raises some questions in terms of the highest MC temperature the approach should start from, number of MC steps needed at each MC temperature, and what is the most suitable MC temperature decrement, among others. 
Thus, we show in Figure~\ref{fig_dE_vs_MCsteps_Nonsequential_All_T100} the energy evolution of the four W-based alloys when applying the MC-DFT method in a non-sequential fashion at a MC temperature of 100K. By non-sequentially we mean that the starting point of this set of simulations is still an SQS (for each W-alloy) with SRO = 0 for its first nearest-neighbor (NN) shell, but we are imposing directly a specific MC temperature of interest and not a sequence of temperatures as it is shown in Fig.~\ref{fig_dE_vs_MCsteps_sequential_All}.

\section{Conclusions} \label{sec_Conclusions}

In summary, this work presents a computational approach that uses first-principles DFT electronic structure calculations to understand the effect of the chemical environment on tungsten-based alloys. In particular, we compared the Special Quasi-random Structure (SQS) and the DFT-coupled Monte Carlo (MC-DFT) methods when investigating the short-range order and elastic properties of two equimolar WCrTaTi and WVTaTi W-HEAs, and two WCrTi and WCrY W-SAs.
We found that structures after MC-DFT calculation have lower cohesive energies than SQS structures, with shorter lattice constants, and large elastic properties values. Furthermore, distinct element segregation was studied by calculating the SRO parameter and radius distribution function. The total density of states suggested that the existence of SRO could improve the stability of the structure.

\section*{Acknowledgments}

This work used the Extreme Science and Engineering Discovery Environment (XSEDE), which is supported by National Science Foundation grant number ACI-1548562. Specifically, YQ and DC acknowledge support from XSEDE allocation MAT200015. YQ and DC also acknowledge computer time allocations at Villanova's Augie and Alipi clusters. DC acknowledges support from the U.S. Department of Energy, Office of Science, Fusion Energy Sciences Program Early Career Research Program under Award Number DE-SC0023072. MRG and DNM acknowledge funding from the EPSRC Energy Programme [Grant number EP/W006839/1]. DNM work has also been carried out within the framework of the EUROfusion Consortium, funded by the European Union via the Euratom Research and Training Programme (Grant agreement no. 101,052,200 EUROfusion). Views and opinions expressed are, however, those of the author(s) only and do not necessarily reflect those of the European Union or the European Commission. Neither the European Union nor the European Commission can be held responsible for them. LD acknowledges support from LabEx DAMAS (program investissements dAvenir ANR-11-LABX-008-01). 




\clearpage



\clearpage
\bibliographystyle{elsarticle-num}
\bibliography{biblio_yichen_paper3}





\end{document}